\documentclass[epj]{webofc}
\usepackage[varg]{txfonts}   

\wocname{EPJ Web of Conferences}
\woctitle{INPC 2013}%
\begin{document}
\title{Pairing properties of the inner crust of neutron stars at finite temperature}
%
%

\author{Alessandro Pastore\inst{1}\fnsep\thanks{\email{apastore@ulb.ac.be}} 
}

\institute{IAA, CP 226, Universit\'e Libre de Bruxelles, B-1050 Brussels,~Belgium 
          }

\abstract{%
We investigate the thermal properties of the inner crust of a neutron star using the Hartree-Fock-Bogoliubov (HFB) formalism at finite temperature.
We compare our results with the ones obtained solving the same equations, but within the BCS approximation.
We observe that for the outermost regions of the inner crust, the two methods can show important differences, in particular when we use them to calculate the neutron specific heat of the system.
}
\maketitle
\section{Introduction}

The thermal properties of the inner crust of a neutron star have important effects on the cooling time of an isolated neutron star ~\cite{lattimer}.
This region  is composed by nuclear clusters surrounded by a sea of unbound neutrons and ultra-relativistic electrons~\cite{chamel}.
A common model, used to describe the crust, relies on the Wigner-Seitz (WS) approximation~\cite{wigner}; $i.e.$  the crust is divided in a set on independent spheres of radius $R_{\text{WS}}$  centered around each nucleus.
A comparison with the full band theory~\cite{chamel2} has shown that the WS approximation is well suited to describe the ground-state properties of the outermost regions of the crust. For  the regions of the crust  closer to the star core, the validity of such approximation is still under debate.

The first microscopic approach to determine the structure of the inner crust has been done by Negele and Vautherin~\cite{nv} in 1973.
Since then, several groups have investigated how different assumptions affect the final result as the choice of the boundary conditions~\cite{baldo},  of the pairing interaction~\cite{fortin,pizzochero} and of symmetry-energy~\cite{chamel3} among others. 
In ref.~\cite{pastoreA}, it has been shown that the pairing properties of the system turn out to be rather independent of the exact ($R_{\text{WS}}$, Z) configurations adopted for the description a given density region.
The aim of the present article is to continue the investigation done in ref.~\cite{pastoreA} to include finite temperature effects and in particular the impact of the BCS approximation~\cite{ring} on the thermal properties of the inner crust.
The latter can be very advantageous from the numerical point of view especially when dealing with finite range interactions.
It is thus very important to check the validity of such approximation for WS cells compared to a full HFB calculation.

\section{Results}

\begin{table}
\begin{center}
\begin{tabular}{cccccccc}
\hline
\hline
Zone & Z & N& $R_{WS}$ [fm] & $\bar{\rho} $ [fm$^{-3}$] & $\rho_n ^{b}$ [fm$^{-3}$] &$k_{F}^{n}$ [fm$^{-1}$]  \\
\hline
11 & 40 & 140 & 53.6 &$2.79\cdot 10^{-4}$ & $7.93\cdot 10^{-5}$ & 0.13\\
10 & 40 & 160 & 49.2 &$4.01\cdot 10^{-4}$  & $1.38\cdot 10^{-4}$ & 0.16\\
 9 &  40 & 210 & 46.4 & $5.97\cdot 10^{-4}$&$2.78\cdot 10^{-4}$ & 0.20 \\
 8 & 40 & 280 & 44.4 &$8.73\cdot 10^{-4}$ &$5.02\cdot 10^{-4}$ & 0.24 \\
 7 &  40 & 460 & 42.2 & $1.59\cdot 10^{-3}$&$1.15\cdot 10^{-3}$ & 0.32 \\
 6 &  50 & 900 & 39.3 & $3.74\cdot 10^{-3}$ &$2.99\cdot 10^{-3}$ & 0.44 \\
 5 &  50 & 1050 & 35.7 &$5.77\cdot 10^{-3}$ &$4.75\cdot10^{-3}$ &0.52 \\
 4 &  50 & 1300 & 33.0 &$8.97\cdot 10^{-3}$ &$7.54\cdot 10^{-3}$  &0.61 \\
 3 &  50 & 1750 & 27.6 &$2.04\cdot 10^{-2}$ &$1.77\cdot 10^{-2}$  &0.81\\
 2 &  40 & 1460 & 19.6 &$4.76\cdot 10^{-2}$ &$4.23\cdot 10^{-2}$  &1.08 \\
\hline
\hline
\end{tabular}
\caption{The WS cells used in the present study. In the different columns we have: the particle numbers Z (protons), N (neutrons), the  radius $R_{WS}$, the total average density of the cell, $\bar{\rho} $, the background neutron density $\rho_n^{b}$~(obtained  averaging the  neutron gas density far away  from the cluster) and its  Fermi momentum $k_{F}^{n}$.}
\end{center}
\label{tabWS}
\end{table}

Following   ref~\cite{nv}, the inner crust of a neutron star has been divided in 10 layers of different density, each one is characterized by a proton  $Z$  and neutron number $N$ and radius $R_{\text{WS}}$. The exact values are reported in Tab.\ref{tabWS}.
To study the effect of pairing correlations  on the thermal properties of the crust,  we solve  the Hartree-Fock-Bogoliubov (HFB) equations at finite temperature~\cite{Goodman86} for each WS cell. We will perform two type of calculations: once  we will solve fully self-consistently the HFB equations and once we will adopt the BCS approximation. 
We refer  to ref.~\cite{pastoreA,pastoreB}  for more a  details discussions on the adopted numerical methods.

The HFB (BCS ) equations are solved  using a Skyrme zero range interaction, SLy4~\cite{chabanat}, for the $ph$-channel, and a Gogny D1 interaction in its separable form~\cite{Tian_2009a,Duguet_2004} for the $pp$-channel.
Higher-order pairing correlations are expected to play a role in WS cells, where the exchange of collective vibrations~\cite{Bar10} lead  to a suppression of the pairing gap~\cite{lombardo}. This is clearly an important effect, but at present there is no clear method on how to include such effects in a consistent way in a mean-filed description.

\begin{figure}[!h]
\includegraphics[width=5.5cm,angle=-90,clip]{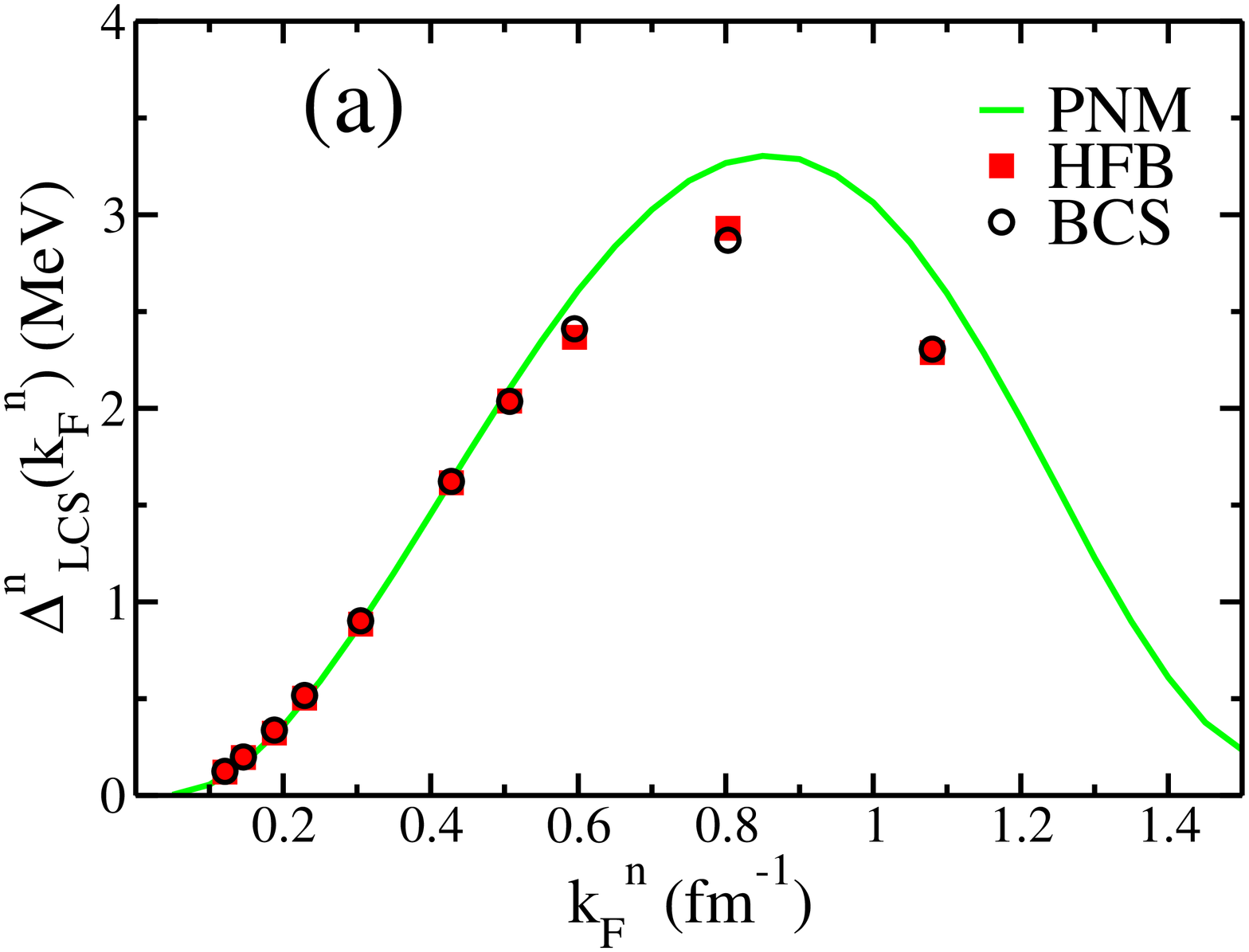}
\includegraphics[width=5.5cm,angle=-90,clip]{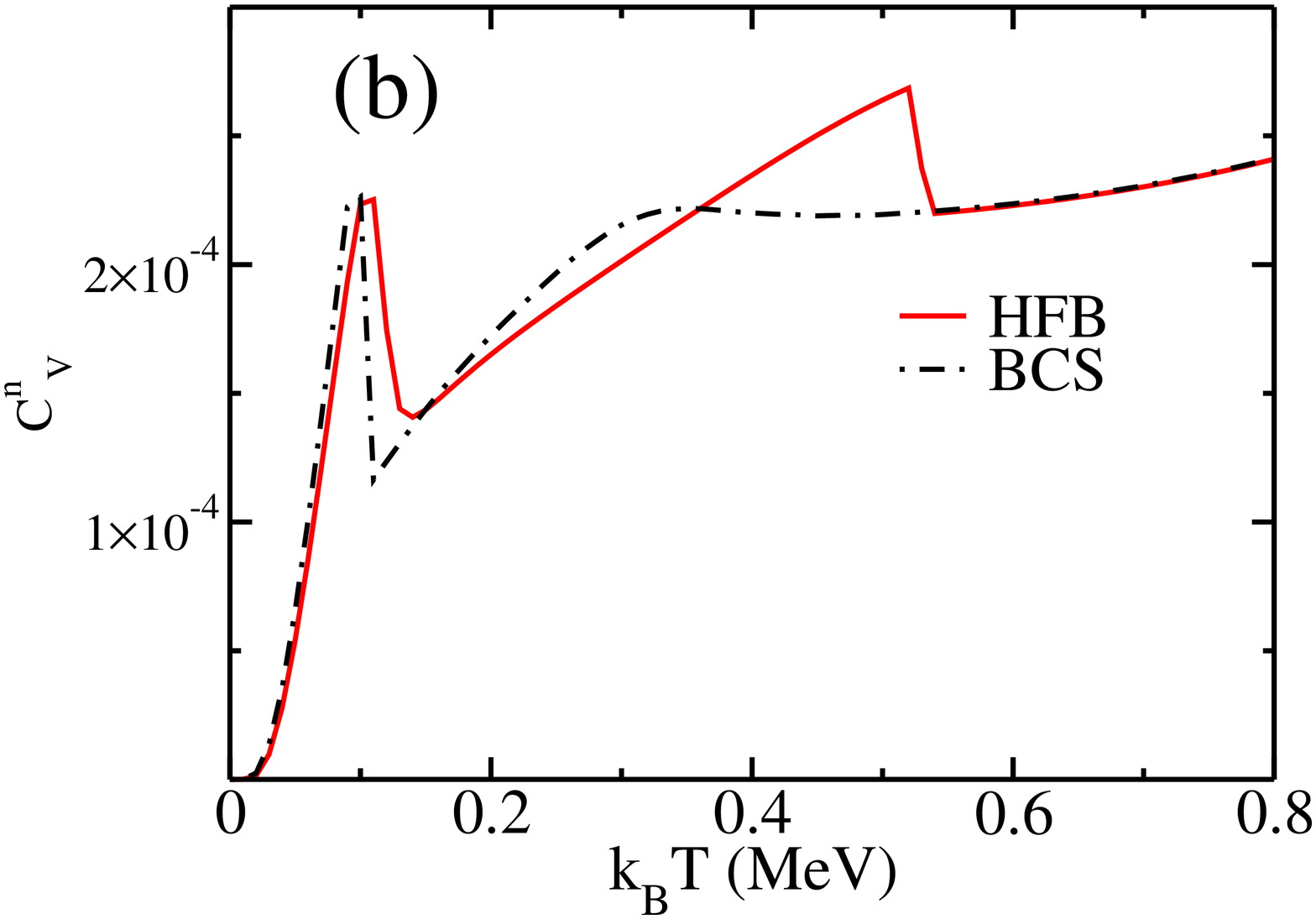}
\caption{(Colors online) Left panel (a): pairing gaps calculated in PNM, solid line, and in the 10 WS cells solving the full HFB equations (squares) and in the BCS approximation (circles). Right panel (b): neutron specific heat, $C^{n}_{V}$, for the cell $^{200}$Zr using the HFB equations (solid line) and the BCS approximation (dash-dotted line). See text for details. }
\label{comp:hfbbcs}       
\end{figure}

In Fig~\ref{comp:hfbbcs} (a) we show the pairing gap at the Fermi energy, $\Delta_{\text{LCS}}^{n}$, at $k_{B}T=0$ for the 10 WS cells of Tab.\ref{tabWS} calculated using the complete HFB equations or the BCS approximation.
Each cell is characterized by a value of the neutron Fermi momentum $k_{F}^{n}=(3\pi^{2}\rho_{n}^{b})^{1/3}$,  calculated using the average neutron density of the free neutron gas far away from the cluster,  $\rho_n ^{b}$, for each cell.
In the same figure we also represent the value of the pairing gap calculated in pure neutron matter (PNM). See ref.~\cite{pastoreA} for more details.
The presence of nucleus at the center of the WS cell, reduces the value of the pairing gap compared to the homogeneous case for high density cells~\cite{pizzochero}, the same result  has been obtained using both HFB and  BCS method.

\begin{figure}[!h]
\includegraphics[width=5.5cm,angle=-90,clip]{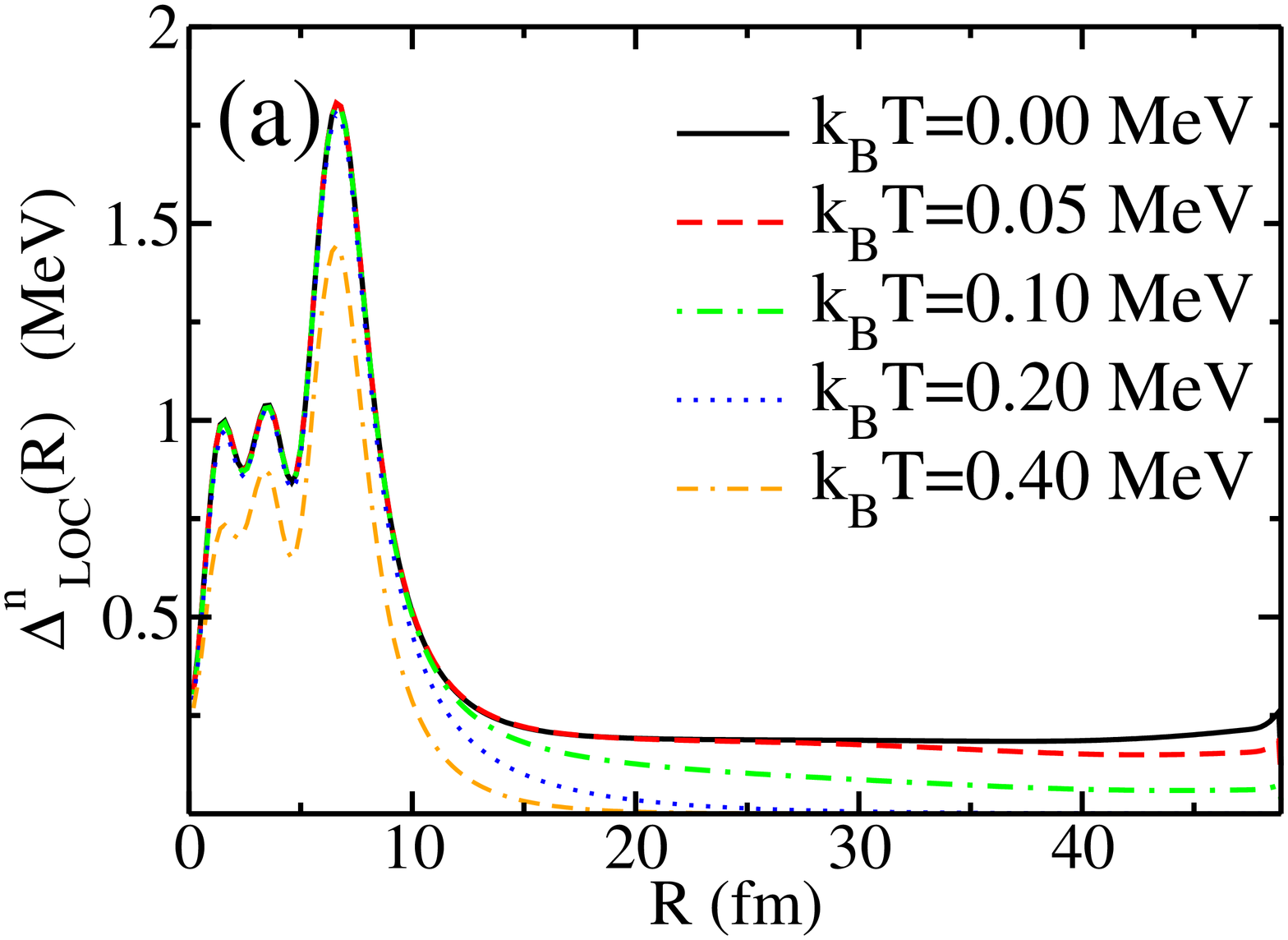}
\includegraphics[width=5.5cm,angle=-90,clip]{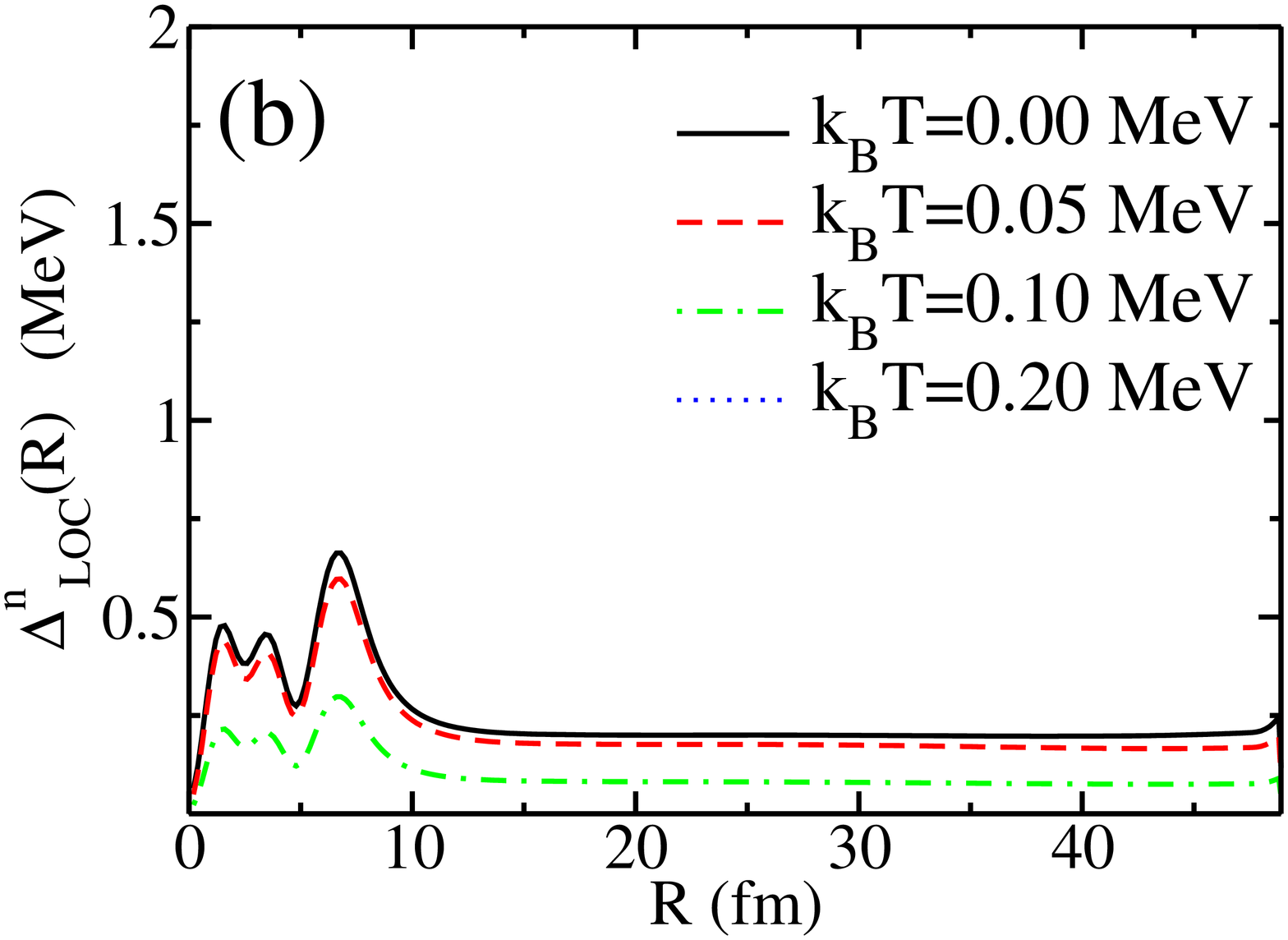}
\caption{(Colors online) Local neutron pairing field, $\Delta_{\text{LOC}}^{n}(R)$, calculated at different temperatures for the cell $^{200}$Zr. On the left panel (a) we show the HFB result, while on the right panel (b) the BCS case. See text for details. }
\label{comp:hfbbcs2}       
\end{figure}

The situation is quite different when we look at  the neutron specific heat, $C^{n}_{V}$.
In Fig.\ref{comp:hfbbcs} (b), we compare this quantity for a low-density cell $^{200}$Zr.
The solid line represents the neutron specific heat calculated using the complete HFB equations, while the dashed line represents the same calculation, but using the BCS approximation.
We observe that the first phase transition happens at $k_{B}T^{1}_{c}=0.1$ MeV. Such value is also compatible with the formula derived from BCS theory~\cite{bcsbook}. 

\begin{equation}
k_{B}T_{c}=\frac{\text{exp}(\zeta)}{\pi}\Delta(T=0),
\end{equation}

\noindent where $\zeta\approx0.577$ is the Euler-Mascheroni constant and  $\Delta(T=0)$ is the pairing gap of the system at zero temperature.
Thus, as already discussed in ref.~\cite{pastoreB}, the first peak corresponds to the disappearance of pairing in the gas.
When we increase the temperature, we observe a second peak at $k_{B}T_{c}^{2}$ in the specific heat, but  for the HFB case only.
This second phase transition has been found also by other authors~\cite{fortin} and it corresponds to the disappearance of the pairing gap inside the nuclear cluster.
To better clarify this difference, in Fig.\ref{comp:hfbbcs2}, we show  the neutron  local pairing field, $\Delta_{\text{LOC}}^{n}(R)$~\cite{pastoreD,pastoreE}, at several temperatures obtained from full HFB and BCS calculations for the cell $^{200}$Zr.
In the HFB case ,when the temperature goes from zero to $T_{c}^{1}$, the neutron pairing field decrease in the gas region, while it does not change in the interior. Going from $T_{c}^{1}$ to $T_{c}^{2}$, we notice that it is now the pairing field inside the region of the nucleus that goes to zero.
In the BCS case, the pairing field does not distinguish among the gas and the nucleus and it presents just one critical temperature.

For the higher density cells, we observe only one peak in the specific heat $C^{n}_{V}$~\cite{pastoreB} and we have found that HFB and BCS are in good agreement.

\section{Conclusions}

In the present article we have analyzed in detail the thermal properties of some WS cell with particular attention to the low-density one as $^{200}$Zr.
The results have been obtained solving both the HFB equations fully self-consistently and  the BCS approximation.
We have found that  the HFB and BCS can present some remarkable differences when dealing with specific quantities in the low-density region of the inner crust of a neutron star.
In particular, only a complete HFB calculation is able to describe the two-peaks structure of the neutron specific heat found in the cell $^{200}$Zr~\cite{pastoreB}.
Such structure in $C^{n}_{V}$ has been also observed in HFB calculations based on density dependent contact pairing interactions~\cite{fortin}, and it has been shown that it is a general feature of low-density WS cells and it is independent from the adopted pairing interaction~\cite{pastoreB}.
The simple BCS approximation predicts for the same system just one phase transition.
As already found in ref.~\cite{pastorec}, the differences between BCS and HFB are large in low density WS cells and they become more and more similar when the gas density increases and the WS cell tends toward a more homogeneous system.

%
%

\end{document}